\documentclass[useAMS]{mn2e}
\usepackage{psfig}

\usepackage{times}

\title[Optical afterglows of gamma--ray bursts: a bimodal distribution?]
{Optical afterglows of gamma--ray bursts: a bimodal distribution?}

\author[Marco Nardini, Gabriele Ghisellini, Giancarlo Ghirlanda]
{Marco Nardini $^{1}$\thanks{E--mail: nardini@sissa.it},
Gabriele Ghisellini $^{2}$ and 
Giancarlo Ghirlanda $^{2}$\\
$^{1}$SISSA--ISAS, Via Beirut 2-4, 34314, Trieste, Italy\\
$^{2}$Osservatorio Astronomico di Brera, via Bianchi 46, I--23807
Merate, Italy.
}
\begin{document}

\maketitle

\label{firstpage}

\begin{abstract}
  
  The luminosities of the optical afterglows of Gamma Ray Bursts, 12 hours
  (rest frame time) after the trigger, show a surprising clustering, with a
  minority of events being at a significant smaller luminosity.  If real, this
  dichotomy would be a crucial clue to understand the nature of optically dark
  afterglows, i.e. bursts that are detected in the X--ray band, but not in the
  optical.  We investigate this issue by studying bursts of the pre--{\it
    Swift} era, both detected and undetected in the optical.  The limiting
  magnitudes of the undetected ones are used to construct the probability that
  a generic bursts is observed down to a given magnitude limit.  Then, by
  simulating a large number of bursts with pre--assigned characteristics, we
  can compare the properties of the observed optical luminosity distribution
  with the simulated one.  Our results suggest that the hints of bimodality
  present in the observed distribution reflects a real bimodality: either the
  optical luminosity distributions of bursts is intrinsically bimodal, or
  there exists a population of bursts with a quite significant grey
  absorption, i.e. wavelength independent extinction.  This population of
  intrinsically weak or grey--absorbed events can be associated to dark
  bursts.
\end{abstract}


\section{Introduction}

Since the detection of the first long Gamma Ray Burst (GRB) optical afterglow
(Van Paradijs et al. 1997), the non--detection of any optical source in the
direction of the gamma--ray trigger of some events stimulated the interest
about the possible differences between the nature of the afterglow emission of
the optically bright and faint GRBs.  During the last 9 years the increasing
number of optical detections and spectroscopic redshift determinations allowed
us to study the intrinsic features of the optical afterglow emission of long
GRBs.  Despite the improved (i.e. made more promptly) observations of optical
afterglows, in almost half of the observed long GRBs no optical counterpart is
still found.  These events have been called in the literature as Dark Burst,
or Failed Optical Afterglows GRBs (Lazzati et al. 2002).

\begin{figure}
\vskip -0.5 true cm
\centerline{\psfig{figure=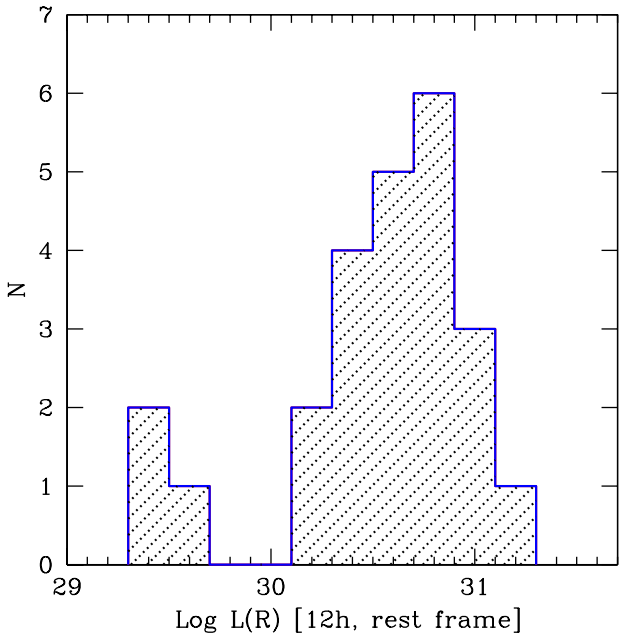,angle=0,width=10cm}}
\vskip -0.8 true cm
\caption{
  Histogram of the monochromatic optical luminosities 12 hours (rest frame)
  after the trigger for the 24 GRBs analysed in Nardini et al. (2006a).  Data
  have been de-reddened both for Galactic and host extinction.  }
\label{istobs}
\end{figure}

In Nardini et al. (2006a) \footnote{Liang \& Zhang (2006) independently found
  similar results.}  we showed the optical $R$ band luminosity light curves of
a sample of 24 pre--{\it Swift} GRBs with known spectroscopic redshift
  and published   
estimate of host galaxy dust absorption.  We found a strong clustering of that
luminosities for the GRBs in our sample.  Most (i.e. 21/24) of the $R$ band
luminosities at 12 hours in the source frame are clustered within a
log--normal distribution centred around a mean value $\log L_{\nu_R}=30.65$
[erg s$^{-1}$ Hz$^{-1}$] with a dispersion $\sigma=0.28$.  We also found 3
GRBs showing dimmer luminosities, a factor 15 (from 3.6 to 4.6
  $\sigma$) smaller than the mean of the 
higher luminosity distribution.  No GRB was found in the luminosity range
between these two ``families''.  In Fig \ref{istobs} we show the histogram of
the $R$ band luminosities of our sample of GRBs.

In a recent update (Nardini et al. 2006b) we added 8 new GRBs detected by
the {\it Swift} satellite (Gehrels et al. 2004) for whose an estimate
of the host galaxy dust extinction has been published.  This small
sample of {\it Swift} GRBs confirms both the clustering and the
bimodality of the optical luminosities found by us with pre--{\it
  Swift} bursts.  We also evaluated the optical luminosities for 17
other {\it Swift} GRBs without any published $A_V^{host}$ estimate,
and found that they are consistent with our previous findings. 

The discovery of a family of optically dim GRBs is an important clue for the
understanding of the nature of dark bursts. The few underluminous observed
events could be the tip of the iceberg of a population of GRBs which
are intrinsically less luminous.  Therefore, a 
fraction of dark GRBs could belong to this family whose distance, optical
absorption, or observing conditions do not allow any optical detection.
Because of its potential importance for the understanding of dark bursts,
  we investigate, in this paper, if the observed bimodal luminosity
  distribution of optically--bright bursts is due to any selection effect
  (related to the search/detection of GRB optical counterparts) or if it
  reflects the existence of two GRB populations.

To this aim we simulate through a Montecarlo method a sample of GRBs with a
redshift distribution traced by the cosmic star formation rate (Porciani \&
Madau 2001), assuming different shapes of their intrinsic optical luminosity
function.  We also simulate different values of dust absorption within the
host galaxy to all the simulated events.  This value is calculated assuming
the standard extinction curves (Pei 1992).

We infer a limiting magnitude distribution obtained by the analysis of
the deepest $R$ band upper limits of all the pre--{\it Swift} GRBs with
no detection of their optical afterglow.  This is the key point of our study:
the use of the upper limits on the optical flux to construct the probability
that a simulated bursts would be detected or not.  It is this probability
distribution that allows us to perform, meaningfully, our simulations.  We
then compare the resulting luminosity distribution of the detectable
simulated events with the one obtained in Nardini et al. (2006a) and
shown in Fig \ref{istobs}.

The scope of our simulation is to check if for any conceivable
  combination of the input assumptions (i.e. luminosity function,
  extinction and redshift distribution) we can reproduce a simulated
  sample whose $R$--band luminosity distribution (12h rest frame) is
  consistent with that observed with the sample of 24 pre--{\it Swift}
  GRBs.

For our analysis we use only the pre--{\it Swift} GRBs because they represent
an homogeneous sample: their optical light curves were sampled from hours
  to days after the trigger and the estimates of the host galaxy extinction
  have been published. 
The same study can be repeated using the more
recent {\it Swift} GRBs once we will have a sufficient number of events with
published estimates of the host galaxy extinction (now this number is too
small, see Nardini et al.  2006b).

\section{Optical upper limits of dark bursts}

During over 7 years from the detection of the first long GRB optical afterglow
in 1997 (Van Paradijs et al. 1997) to the launch of the {\it Swift}
satellite on 
November 20th 2004 (Gherels et al. 2004), 238 long GRBs have been localised,
within a few hours to days, with an accuracy of 1 degree or better.  For only
64 of them an associated optical afterglow was found.  For a large fraction of
them the lack of optical data is due to the absence of any optical telescope
pointing the source location error.  We found 111 bursts with at least one
optical--NIR failed (i.e.  giving only a flux upper limit) observation and
among them we focused our attention on the 94 GRBs with at least one $R$ band
limiting magnitude.

In order to avoid including in our sample events with a large gamma--ray error
box uncovered by the optical observation, we discarded the events with an
error box wider than 11' of radius if the $R$ band observation set does not
cover at least the 80\% of the error box area.  We found 2 events which do not
satisfy this criterion.  For the others:
\begin{itemize}
\item 58 events have an error box narrower than 10' of radius
\item 27 events with the entire error box covered by the observations
\item 4 events with more than 90\% of the error box covered
\item 3 events with more than 80\% of the error box covered
\end{itemize}
We performed our analysis with these 92 dark GRBs.

We often found a large number of $R$ band upper limits for a single burst
obtained at different epochs after the trigger. In these cases we evaluated
the deepest limiting $R$ band magnitude for each GRB assuming a temporal
behaviour $F(\nu , t)\propto t^{-\alpha}$ with $\alpha=1$, the average slope
of the detected optical afterglows. 
For instance suppose that, for a given burst, there are two upper
limits of $R>18$ and $R>20$ at, say, 1 and 24 hours since trigger,
respectively.  
We select $R>18$ at 1 hour as the most stringent, since it corresponds
[assuming $F(t)\propto t^{-1}$] to $R>21.45$ at 24 hours.

We corrected the resulting upper limits for the Galactic dust extinction along
the line of sight using the absorption maps found by Schlegel et al. 1998.
For most of the events the amount of Galactic dust absorption is negligible
but there are some GRBs absorbed by several magnitudes in the $R$ band .  For
example, along the line of sight of GRB 030501 and GRB 030320, the Galactic
dust absorption value $A_R$ (in the $R$ band) is 39 and 20.5 magnitudes,
respectively.  Such an extinction makes impossible any GRB optical afterglow
detection.
    
\begin{figure}
\vskip -0.5 true cm
\centerline{\psfig{figure=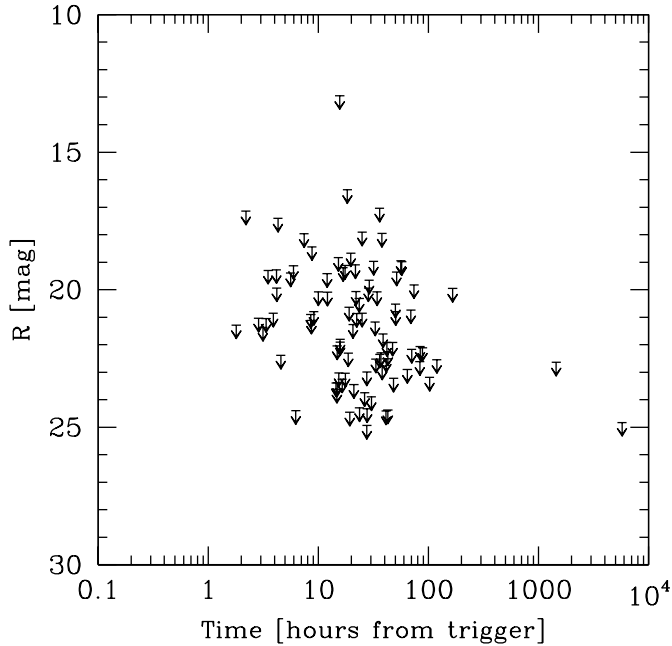,angle=0,width=10cm}}
\vskip -0.8 true cm
\caption{Deepest $R$ band upper limits for all the pre--{\it Swift}
  dark bursts. All the data are corrected for the Galactic extinction
  given in Schlegel et 
  al. (1998). Each upper limit corresponds to a single GRB. For GRBs
  with several upper limits available in the literature, we report,
  conservatively, the deepest upper limit evaluated by assuming a
  standard flux decay light curve (see text). }
\label{ul_min}
\end{figure}

In Fig. \ref{ul_min} we show the deepest $R$ band upper limits for all the 92
GRBs of the sample, de-reddened for the Milky Way dust absorption\footnote{The
  references for the upper limit values are reported in the appendix.}.

\subsection{Telescope Selection Function}

For the majority of the dark GRBs in our sample, the deepest upper limit is
quite constraining. Lazzati et al. (2002), albeit with a smaller
sample of bright and dark GRBs, demonstrated that the non detection
of an optical afterglow for most of the optically dark GRBs was not
due to adverse observing conditions or delay in performing the
observations. They also showed that these events do not have particularly
large Galactic absorbing columns. The upper limits we added in our
sample are generally deeper than the previous ones, thus
confirming their results.   
Therefore, we can reasonably exclude that any instrumental bias is
responsible for the failure of the detection of the optical afterglow
in dark GRBs.    

In order to obtain an homogeneous distribution of upper limits we extrapolated
the deepest limiting $R$ band magnitude for each burst at the common time of
12 hours after the burst trigger (observer frame).  We again assumed a
temporal behaviour of the form $F\propto t^{-\alpha}$ with an index
$\alpha=1$.

\begin{figure}
\vskip -0.5 true cm
\centerline{\psfig{figure=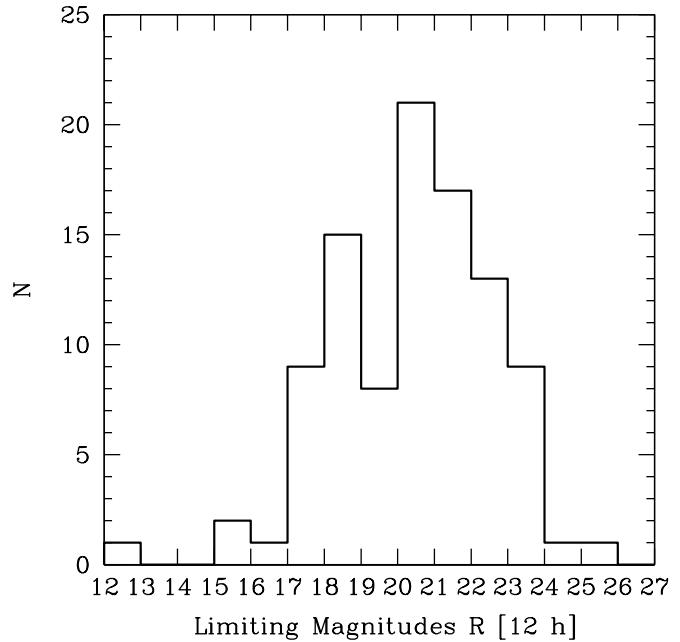,angle=0,width=10cm}}
\vskip -0.8 true cm
\caption{
Histogram of the deepest $R$ band upper limits
(corrected for Galactic extinction) extrapolated at a common time (12h after
the trigger), 
obtained assuming a temporal behaviour $F(t)\propto t^{-1}$
of the optical afterglow flux. This distribution
  corresponds to the Telescope Selection Function (see text).
}  
\label{istoul}
\end{figure}

In Fig. \ref{istoul} we show the $R$ band deepest limiting magnitudes for all
the dark GRBs in our sample.  The obtained values represent the distribution
of the optical observation depth for a large number of events.  
Thanks to the consistency of the upper limits and the afterglow
detections (see Lazzati et al. 2002),  
we can use this distribution to describe the probability for a burst
to be observed in the optical with a certain depth.  We can call this
distribution \emph{Telescope Selection Function} (\textbf{TSF}).

We can see in Fig.~\ref{istoul} that most of the dark bursts have been
observed at 12 hours at least down to R$\approx$20.
Sometimes the limiting magnitudes are greater than 24.  Only for a small
fraction of GRBs the limiting magnitude is smaller than 15, possibly
due to bad observational conditions and/or high Galactic absorption.

Note that all the limiting magnitudes and the detected afterglow data
considered in our sample concerns bursts observed before the launch of the
{\it Swift} satellite.  During the last two years, indeed, the prompt
(within few minutes since trigger) location of the GRB allowed the
early pointing of the optical telescopes. As a consequence, dark
bursts observed in the Swift--era have systematically earlier upper
limits than pre--Swift bursts.  
Besides, the early optical (and at a
larger extent the early X--ray) 
  light curves of several Swift bursts have shown unexpected features
  (different slopes, re-brightening and flares) whose nature is still debated.
  Moreover, the increased number of GRBs with accurate and promptly
  distributed positions makes it difficult to systematically extend the
  optical follow--up campaign up to few days after the trigger for all bursts.
  Therefore, on average, the available optical observations of Swift bursts
  are covering the very early optical afterglow emission, from few minutes to
  several hours since the burst onset.
For these differences we prefer to keep separate the pre--Swift and the
  Swift bursts although the study that we propose in this paper, based on the
  sample of pre--Swift GRBs, can be performed with a sizable sample of Swift
  bursts with host extinction and late optical observations.

\section{Simulated sample}

The basic idea of our simulation is to produce, under some assumptions, a
  population of GRBs which is ``subject'' to the same TSF that we constructed
  from the upper limits of dark bursts. The result is a population of observed
  optical afterglows which can be compared with the real one. This is a
  test on the assumptions, of the simulated sample, i.e. (1) its redshift
  distribution, (2) the intrinsic luminosity function and (3) the host galaxy
  extinction.

\subsection{Redshift distribution}

The lack of optical information about the dark GRBs does not allow a direct
spectroscopic redshift determination for all but two of them (GRB 000210, Piro
et al 2002; GRB 000214, Antonelli et al. 2000).  Assuming that the dark GRBs
are related to the same progenitors of the optically detected bursts we can
use the cosmic star formation (CSFR) history to represent the redshift
distribution of all the GRBs we analyse.
Among the three recipes of Porciani \& Madau (2001), which differ at $z\ge$2,
we considered  the CSFR\#2 (Eq.5 in that paper).  The K--correction
has been calculated assuming that the optical--UV afterglow spectrum is a
single power law: $F(\nu)\propto \nu^{-\beta}$.  The observed spectral index
$\beta$ is usually in the range $0.5<\beta <1$.  We used a typical value
$\beta=1$ in our simulation but our results are unchanged if adopting
different values in the range $0.5<\beta<1$ and choosing a different
shape for the CSFR (e.g. Eq. 4 or 6 in Porciani \& Madau (2001)).

\subsection{Luminosity function}

We assumed 3 different types of intrinsic monochromatic luminosity
distributions at the common time 12 hours in the source frame: a
log--normal, a powerlaw and a top--hat. For each luminosity function
we considered several combination of their free parameters. Note that
we cannot be guided, in the choice of the optical luminosity 
function, from what we already know of the luminosity function of the
prompt emission of GRBs (see e.g. Firmani et al. 2004), since there 
is no correlation between the optical luminosity at 12 hours and 
the luminosity (or total energy) of the burst (see Nardini et al. 2006).

\subsection{Host galaxy dust absorption}

The association of long GRBs with massive progenitors could imply the presence
of a large amount of absorbing dust in the source neighbourhood.  On the other
hand, the analysis of optical--near infrared afterglow spectral energy
 distributions showed a relatively small amount of reddening due to dust in the
host galaxy.  The value of $A_V^{host}$ in the source frame is usually of the
order of a fraction of a magnitude (Kann et al. 2006; see Fig. 3 in Nardini et
al. 2006a), despite the evidence of high $N_{\rm H}$ column densities found in
the X--ray afterglow analysis (Vreeswijk et al. 2004; Jakobsson et al. 2006;
Stratta et al. 2004).

We take into account the host galaxy dust absorption effects on the observable
optical luminosities in our simulation.  We test different shapes of the
intrinsic $A_V^{host}$ distributions.  The corresponding $A_R^{host}$ (in the
rest frame) has been evaluated using the analytical extinction curves by Pei
(1992).  Most of the estimated dust absorption in optical GRB afterglows are
well described by extinction curves without an evident 2175 \AA~feature, so we
used a Small Magellanic Cloud like extinction curve.  In any case, our results
are not largely affected by this choice.

This sample of generated events is then assumed to be observed using optical
telescopes with a limiting magnitudes distribution traced by the TSF at the
common ($z=0$) time $t_{obs}=12h$.  (Note that the effect of Galactic dust
absorption is considered within the TSF definition).  All events whose
redshift makes the Ly--$\alpha$ break to obscure the observed $R$ band
radiation have been considered as dark. In summary:
\begin{itemize}
\item we assume a redshift distribution function, a luminosity distribution
  and a host galaxy absorption function;
\item we pick up at random a redshift $z$ a luminosity $L(R)$ and a host
  extinction $A_V$. With these parameters we compute the $R(12h)$
  magnitude of the event at 12h in the observer frame;
\item we pick up at random a limiting magnitude $R_{lim}(12h)$ for the
  telescope that will observe this event within the TSF.
\item We compare $R(12h)$ with $R_{lim}(12h)$ to decide if this event can be
  observed or not. All events with redshift greater than 5 are considered
  undetectable.
\end{itemize}
In order to make a statistically meaningful simulation we repeat the above
procedure 1000 times and build up the luminosity distribution of the
``observable'' events.  This distribution can be finally compared with the
observed one (Fig. \ref{istobs}). Through the comparison between the
  simulated ``detectable'' sample and the really observed one we can assign a
  probability to our set of assumptions.  We can then repeat this procedure by
  changing the starting assumptions (e.g. the luminosity and/or absorption
  distribution).

\begin{figure}
\vskip -0.5 true cm
\centerline{\psfig{figure=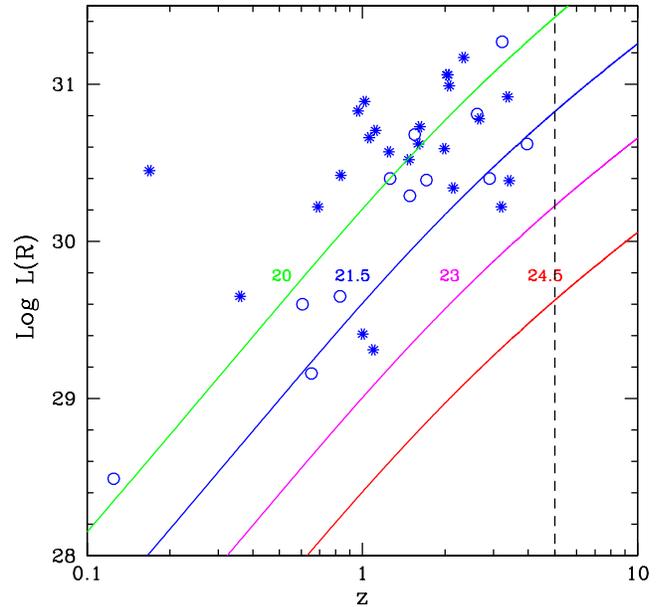,angle=0,width=10cm}}
\vskip -0.8 true cm
\caption{ Limiting observable intrinsic $R$ band luminosity as a
  function of redshift, for different limiting magnitudes. 
Starred dots represent the luminosities of the GRBs in the sample and
  circles represent the Swift burst luminosities (Here we added 4 new
  Swift GRBs with a published estimate of $A_V^{host}$). 
The dashed vertical line at $z=5$ corresponds the Lyman--$\alpha$ break
for the $R$ band.}
\label{lumlim}
\end{figure}

\section{Comparison with the observed distribution}

The luminosity distribution of the simulated GRBs, that are observable using
the considered TSF, has to be compared with the distribution of the observed
GRB afterglows.  Some features of the tested intrinsic luminosity functions
have been chosen in order to better reproduce the distribution represented in
Fig. \ref{istobs}.  For example it is necessary to impose an high luminosity
cut off to the luminosity function at about $\log[L(\nu_R)^{12h}]\approx 31.2$
[erg s$^{-1}$Hz$^{-1}$].  A GRB with a greater luminosity would be easily
detectable also with a low limiting magnitude for almost all the redshifts
smaller than 5 (see Fig. \ref{lumlim}).  The absence of any observed GRB with
such a luminosity therefore sets a constraint to the luminosity
function. \\
The method most commonly adopted for comparing two distinct distributions is
the two-sample Kolmogorov Smirnov (K-S) test.  Unfortunately, given the
specific luminosity distribution we are considering (Fig \ref{istobs}), this
method has some critical limitations. Indeed (e.g. Press et al. 1992,
Numerical Recipes in C, Second Edition; Ashman et al. 1994), the K-S test is
ideal for comparing the median of two distributions but it is not sensitive to
the tails of the distributions being compared and it also fails in comparing
bimodal with unimodal distributions. In particular in our case we want to
statistically verify a bimodality represented by an excess of events (i.e. the
low luminosity events) which are located on the tail of the more numerous
population of high luminosity bursts but at more than 3.6 $\sigma$ away from
its central value.  By applying the K-S test in order to compare an
  unimodal luminosity function with such an observed distribution we expect to
  strongly overestimate the null hypothesis probability.  This is the reason
why we have searched for a statistical method that fully exploited the
observational constrain of having ``a lack'' of events with optical
luminosities in between those of the two population of Fig.\ref{istobs}.
\\
We adopted the Likelihood Ratio Test (LRT) described by Cash
  1979. \\
  The simulation generates a sample of 30000 GRBs, each with an associated
  intrinsic luminosity $L(\nu_R)$, redshift, host galaxy dust absorption
  $A_V^{host}$ and telescope limiting magnitude. The simulation returns the
  luminosities of the events that have an observer frame flux large enough to
  be detected by the associated telescope. Through this luminosities
  distribution we can predict the number of bursts expected in each bin using
  the same binning adopted in the histogram plotted in Fig.  \ref{istobs}. We
  can then compare these predictions with the observed data by evaluating the
  factor C (eq.3 of Cash 1979) of the LRT. A large Cash statistic C implies
  the rejection of the model. As a reference value we adopted C=9.2 which
  corresponds to a probability of rejection of the model hypothesis of $P_{rej}=90$\%.
  A simulated distribution can be considered consistent  with the
  observed data with $P_{rej}<90$\% if the obtained C is smaller than 9.2.

\begin{table}
\begin{center}
\caption{Simulation results without considering host galaxy absorption}
\begin{tabular}{l l c c c c c c c c}
\hline
\hline
a&          b  &$C^c$  & $CP^d$ & $P_{KS}^e$\\
    &                &    &   &   \\
\hline 
G             &30.65, 0.25  &35.5 & $<10^{-5}$&0.38 \\
G             &30.20, 0.70  &14.8 & $6.1\cdot 10^{-4}$& 0.31\\   
G             &30.50, 0.50  &15.4 & $4.5\cdot 10^{-4}$& 0.69 \\  
\hline  
TH            &29.3, 31.2   &12.0 & $2.5\cdot 10^{-3}$& 0.33\\ 
\hline
PL            &29.3, 31.2, $-$1   &11.7& $2.9\cdot 10^{-3}$ & 0.24\\
PL            &29.3, 31.2, $-$1.5 &11.7& $2.9\cdot 10^{-3}$ & 0.24\\
PL            &29.3, 31.2, $-$2   &11.6& $3.0\cdot 10^{-3}$ & 0.22\\
\hline
\label{noabs}
\end{tabular}
\end{center}
a) Assumed luminosity distribution: G=Gaussian, TH=top hat,
PL=power--law. \\ 
b) Parameters G: $\mu$, $\sigma$; TH: minimum luminosity, maximum
luminosity; PL: minimum luminosity, maximum luminosity, index $\alpha$
assuming $N\propto L(\nu_R)^{\alpha}.$\\ 
c) Value of the C factor obtained with the Cash statistics. \\
d) Cash Probability from eq. 3 (Cash 1979).
 \\  
e) Probability obtained with the K-S test.

\end{table}  

\section{Results}

\subsection{Simulation without considering host galaxy dust absorption}

We considered different combinations of the parameters characterising the
assumed luminosity distribution (i.e. mean value $\mu$ and $\sigma$ for the
log--normal, luminosity range and slope $\alpha$ for the power--law and the
luminosity range for the top hat).  The results of the simulations are listed
in Tab.  \ref{noabs} We found that in none of these cases the observed
luminosity distribution of the simulated samples agrees with the observed one.
The factor C is always larger than 11.6. In the log--normal
  cases, a narrow luminosity function that well matches the observed high
  luminosity peak returns a large C value because it cannot reproduce the low
  luminosity excess.  A too wide distribution has, instead, an excess of
  events with $\log{L(\nu_R)^{12h}} > 31.2$, in contrast with the observed 
  maximum luminosity.

Both the power--law and the top hat distributions are affected by similar
problems.  The observed high luminosity cut off requires an upper bound to the
simulated distributions.  The low luminosity end instead does not affect our
results.  As it happens for the log--normal distribution, we are unable to
reproduce an observable luminosity distribution, since we always violate some
of the observed properties of the distribution shown in Fig \ref{istobs}.

\begin{figure}
\vskip -0.5 true cm
\centerline{\psfig{figure=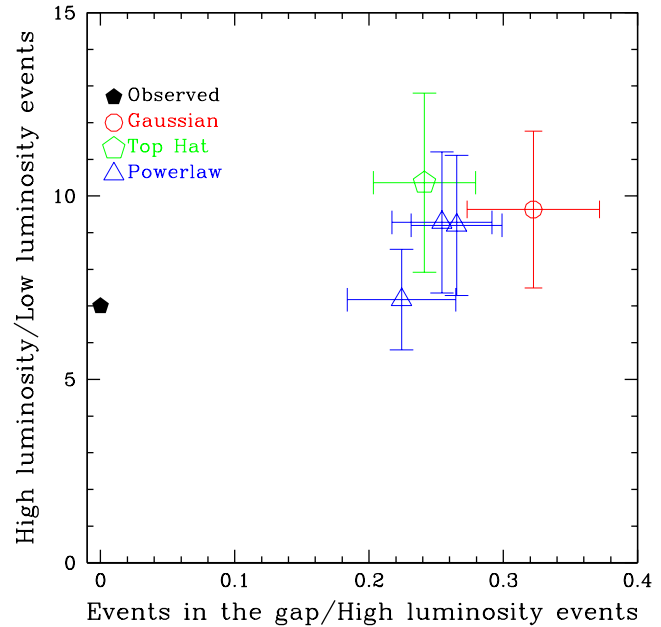,angle=0,width=10cm}}
\vskip -0.8 true cm
\caption{
Unabsorbed case.
Ratio between the number of observed events with
$30.2<\log[L(\nu_R)^{12h}]< 31.2$ and $\log{L(\nu_R)^{12h}}<29.7$
versus the observed events with $29.7 <\log{L(\nu_R)^{12h}}< 30.2$
and $30.2<\log[L(\nu_R)^{12h}]< 31.2$ for the considered initial
luminosity functions in case of host galaxy dust absorption
absence. Error bars show 1 $\sigma$ uncertainties.  
 }
\label{sigmanoabs}
\end{figure}

\bigskip As an illustrative exercise we show in Fig. \ref{sigmanoabs}
  the results obtained by repeating 1000 times the former simulation with 1000
  events. We plot the ratio between the number of observed events in the high
  ($30.2<\log[L(\nu_R)^{12h}]< 31.2$ over the low ($29.7
  <\log{L(\nu_R)^{12h}}< 30.2$) luminosity bins versus the ratio of the number
  of observed events in the gap ($29.7 <\log{L(\nu_R)^{12h}}< 30.2$) and those
  in the high luminosity range ($30.2<\log[L(\nu_R)^{12h}]< 31.2$). Note that
  the the ratios of the simulated samples stand at more than 3$\sigma$ from
  the observed one (filled pentagon).

\begin{figure}
\vskip -0.5 true cm
\centerline{\psfig{figure=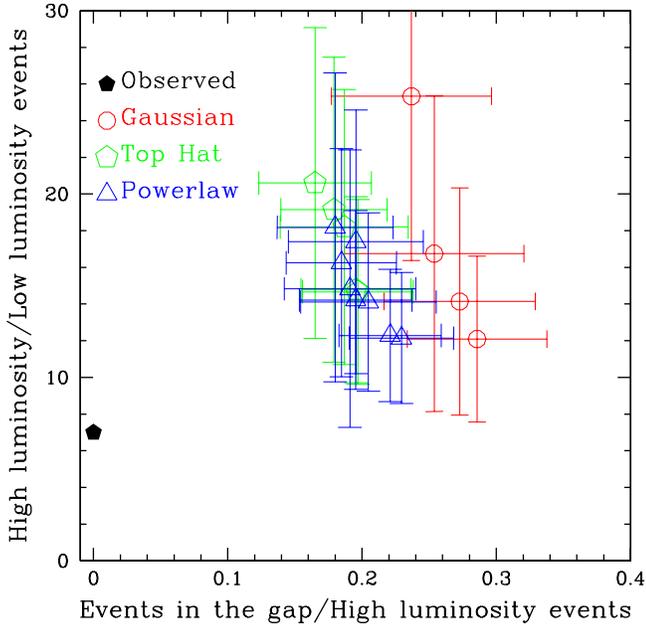,angle=0,width=10cm}}
\vskip -0.8 true cm
\caption{Host galaxy absorption case. Ratio between the
  number of observed events with 
  $30.2<\log[L(\nu_R)^{12h}]< 31.2$ and $\log{L(\nu_R)^{12h}}<29.7$ versus the
  observed events with $29.7 <\log{L(\nu_R)^{12h}}< 30.2$ and
  $30.2<\log[L(\nu_R)^{12h}]< 31.2$ for the different initial luminosity
  functions considering the host galaxy dust absorption effects. Error
  bars show 1 $\sigma$ uncertainties. } 
\label{sigmaabs}
\end{figure}
\subsection{Host galaxy absorption}

In order to check if the addition of the host galaxy dust absorption allows us
to produce a luminosity distribution compatible with the observed one, we
tested some different kinds of absorption distributions and extinction curves.
The distribution of $A_V^{host}$ estimated for the observed bursts is
dominated by low values.  A large number of events are consistent with zero
absorption and the majority of them show $A_V^{host}$ smaller than 1
magnitude.  This however could be due to selection effects, since it is more
difficult to detect highly absorbed optical sources.

We first assumed a top hat $A_V^{host}$ distribution, with a minimum
$A_V^{host}=0$, and assuming different maximum absorption values (i.e.
$A_{V,Max}^{host}=2, 3, 5$ magnitudes).  Then we simulated a power--law like
$A_V^{host}$ distribution (with different slopes and $A_{V,Max}^{host}$),
which better represents the observed distribution.  We finally assumed a
single value for the $V$ band extinction in the host galaxy for all bursts
(note that this conditions in any case implies different values of
$A_{\nu_R(1+z)}$).  We also tried a Gaussian distribution, but the results are
very similar to those found with the top hat distribution.

For all these attempts we have combined the $A_V^{host}$ distributions with
the different luminosity distributions described in the previous section.  The
results are listed in Tab. \ref{siabs}.  In no case we were able to reproduce
the observed distribution.
We conclude that a continuous absorption distribution, combined with a 
unimodal luminosity function, is unable to generate an observable GRBs 
luminosity distribution characterised by an empty gap between the two 
different luminosity groups.  

\begin{table*}
\caption{Simulation results considering different shapes of host
  galaxy dust absorption distributions.}
\begin{center}
\begin{tabular}{l l l c c c c}
\hline
\hline
Luminosity$^a$&Parameters$^a$&Absorption$^b$&$C^a$&$CP^a$&$P_{KS}^a$\\
distribution  &              &parameters    &  & &          \\
\hline 
G             &30.65, 0.25 &TH, 2      &32.3 &$<10^{-5}$& 0.16 \\ 
G             &30.20, 0.70 &TH, 2      &14.4 & $7.5\cdot 10^{-4}$ & 0.20\\ 
G             &30.50, 0.50 &TH, 2      &17.1 & $1.9\cdot 10^{-4}$ & 0.23 \\ 
G             &30.65, 0.25 &P, 2, $-$1 &35.4 & $<10^{-5}$&0.25\\
G             &30.65, 0.25 &P, 2, $-$2 &63.8 & $<10^{-5}$&0.28\\
G             &30.20, 0.70 &P, 2, $-$1 &14.3 & $7.8\cdot 10^{-4}$ & 0.49\\ 
G             &30.20, 0.70 &P, 2, $-$2 &14.7 & $6.4\cdot 10^{-4}$ & 0.35\\
G             &30.50, 0.50 &P, 2, $-$1 &15.3 & $4.8\cdot 10^{-4}$ & 0.50\\ 
G             &30.50, 0.50 &P, 2, $-$2 &14.6 & $6.8\cdot 10^{-4}$ & 0.61 \\
G             &30.65, 0.28 &C, 0.5     &34.2 & $<10^{-5}$&0.16\\
G             &30.20, 0.70 &C, 0.5     &13.9 & $9.6\cdot 10^{-4}$ & 0.23\\
G             &30.50, 0.50 &C, 0.5     &18.2 & $1.1\cdot 10^{-4}$ & 0.24\\
G             &30.65, 0.28 &C, 0.7     &33.3 & $<10^{-5}$&0.14\\
G             &30.20, 0.70 &C, 0.7     &14.2 & $8.3\cdot 10^{-4}$ & 0.14\\
G             &30.50, 0.50 &C, 0.7     &17.5 & $1.6\cdot 10^{-4}$ & 0.17\\
\hline
TH            &29.3, 31.2   &TH, 2     &11.6 & $3.0\cdot 10^{-3}$ & 0.56\\ 
TH            &29.3, 31.2   &P,2, $-$1 &12.1 & $2.4\cdot 10^{-3}$ & 0.48\\ 
TH            &29.3, 31.2   &P,2, $-$2 &12.3 & $2.1\cdot 10^{-3}$ & 0.35\\ 
TH            &29.3, 31.2   &C, 0.5    &11.8 & $2.7\cdot 10^{-3}$ & 0.58\\    
TH            &29.3, 31.2   &C, 0.7    &12.6 & $1.8\cdot 10^{-3}$ & 0.44\\
\hline
PL           &29.3, 31.2, $-$1 &TH, 2       &11.0& $4.1\cdot 10^{-3}$ &0.75\\
PL           &29.3, 31.2, $-$1 &P,  2, $-$1 &11.5& $3.2\cdot 10^{-3}$ &0.41\\ 
PL           &29.3, 31.2, $-$1 &C,  0.7     &10.4& $4.5\cdot 10^{-3}$ &0.64\\
PL           &29.3, 31.2, $-$1 &C,  0.5     &11.3& $3.5\cdot 10^{-3}$ &0.73\\
PL           &29.3, 31.2, $-$2 &TH, 2       &10.6& $5.0\cdot 10^{-3}$ &0.77\\
PL           &29.3, 31.2, $-$2 &P,  2, $-1$ &11.2& $3.7\cdot 10^{-3}$ &0.77\\ 
PL           &29.3, 31.2, $-$2 &C,  0.5     &10.9& $4.3\cdot 10^{-3}$ &0.78\\
PL           &29.3, 31.2, $-$2 &C,  0.7     &10.7& $4.7\cdot 10^{-3}$ &0.71\\
\hline
\label{siabs}
\end{tabular}
\end{center}
a) Same notes as in Tab. \ref{noabs}.\\
b) Absorption distributions parameters. TH: maximum absorption in the $V$
band magnitudes (the minimum is set to 0); P: maximum absorption in the $V$
band magnitudes and index $\alpha$; C: constant absorption $A_V^{host}$.   
\end{table*}  
\begin{table*}
\begin{center}
\caption{Simulation results assuming an achromatic ``grey dust'' absorption}
\begin{tabular}{l l l c c c c c c}
\hline
\hline
Luminosity$^a$&Parameters$^a$&Absorption$^a$&Grey dust$^b$ & $A_{\lambda}$$^c$
&$C^a$&$CP^a$&$P_{KS}^a$\\   
distribution  &            &parameters & \%        &
& &   &     \\ 
\hline 
G         & 30.65, 0.25 &0 &60 &1.6  &6.0 &  $5.0\cdot 10^{-2}$&0.40 \\
G         & 30.65, 0.25 &0 &70 &1.6  &4.4 & 0.11 & 0.69\\
\hline
TH        &30.2, 31.2   &0 &60 &1.5 &5.6 & $6.1\cdot 10^{-2}$ & 0.35\\ 
TH        &30.2, 31.2   &0 &70 &1.5 &4.7 & $9.5\cdot 10^{-2}$ & 0.21\\
\hline
PL        &30.2, 31.2, $-$1 &0 &70 &1.5 &2.9 & 0.24 & 0.50\\
PL        &30.2, 31.2, $-$2 &0 &70 &1.5 &2.6 & 0.27 & 0.51\\
\hline
\hline
\end{tabular}
\end{center}
a) Same notes as in Tab. \ref{siabs}.
\\
b) Fraction of the simulated events with an associated achromatic rest
frame absorption (in percentage).
\\
c) Achromatic absorption amount (in magnitudes).
\end{table*}  

\subsection{Achromatic extinction}

The analysis of the optical to X--ray spectral energy distributions (SED) of
some long GRBs suggests the presence of an achromatic optical absorption
component (Stratta et al. 2005), perhaps due to the small size grain
destruction in the neighbourhood of the GRB (Lazzati et al. 2001, Perna \&
Lazzati 2002).  The amount of this absorption could be higher than what
inferred assuming standard dust (even by several magnitudes).  We then
considered the possibility that a fraction of GRBs can be absorbed with an
achromatic extinction curve.  Starting with the same luminosity function
tested in the previous cases, we associated a further achromatic absorption to
a fraction of events.  Such an absorption decreases the observable flux and
the chance for those events to be detected.  When observed, the analysis of
the optical SED of these GRBs would not show any evidence of dust absorption.
The inferred intrinsic luminosity could therefore be underestimated by a
factor equal to the grey absorption amount.  The optical luminosity
distribution inferred by the observer could appear bimodal even if the real
intrinsic luminosity function were unimodal.
\\
This last absorption model, when applied in the simulation to a large
  fraction of events, returns often a luminosity distribution compatible with
  the observed one. For the cases tested in this paper the C factor is always
  smaller than 9.2. Indeed it is even smaller than 4.6 (P=90\%) reaching in
  some cases very small values with P$<$24\%. We conclude that unimodal
  luminosity functions can reproduce the observed bimodal luminosity
  distribution of fig. \ref{istobs}, if strong achromatic absorption is
  assumed.

\begin{figure}
\vskip -0.5 true cm
\centerline{\psfig{figure=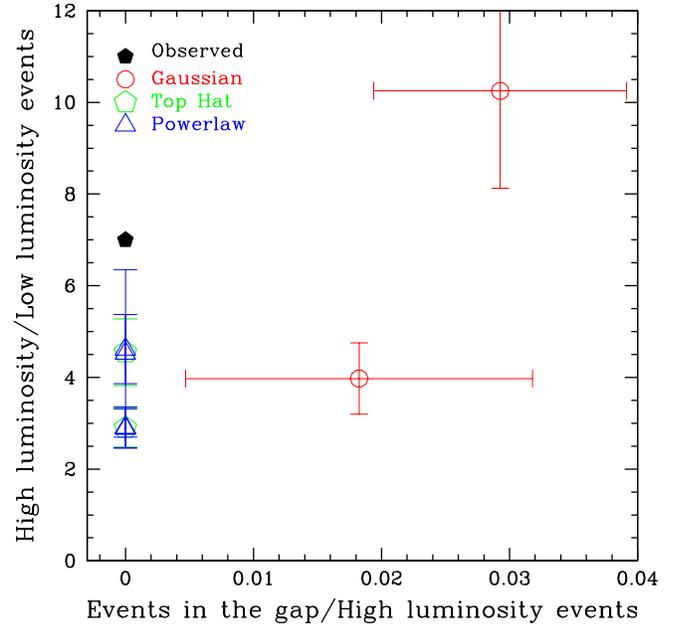,angle=0,width=10cm}}
\vskip -0.8 true cm
\caption{
Achromatic absorption case. Ratio between the number of
observed events with 
$30.2<\log[L(\nu_R)^{12h}]< 31.2$ and $\log{L(\nu_R)^{12h}}<29.7$
versus the observed events with $29.7 <\log{L(\nu_R)^{12h}}< 30.2$
and $30.2<\log[L(\nu_R)^{12h}]< 31.2$ for the different initial
luminosity functions in the case of achromatic dust absorption.  In all
the plotted points we did not consider the contribution of the
standard dust absorption. Error bars show 1 $\sigma$ uncertainties.  
 }
\label{sigmagrigio}
\end{figure}

Similarly to what we have done in Fig. \ref{sigmanoabs} we plot in
  Fig. \ref{sigmagrigio} the high to low luminosity number ratio vs the gap to
  high luminosity number ratio. For clarity, we inserted in this plot only the
  cases with grey absorption only, without the addition of a standard dust
  absorption effect.  As can be seen, there are luminosity functions (powerlaw
  and top--hat) which agree with the observed luminosity distribution.

\section{Discussion}

Our findings suggests that the bimodality of the observed optical luminosity
distribution, even if found with a relatively small number of events, is
significant.  It can be the result of either a bimodality in the optical
luminosity function, or the result of a fraction of bursts being absorbed by a
significant amount of grey dust.

\begin{figure}
\vskip -0.5 true cm
\centerline{\psfig{figure=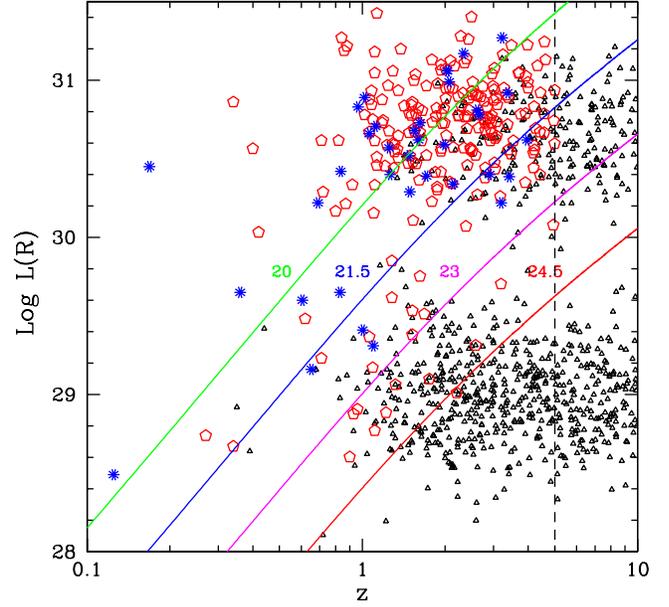,angle=0,width=10cm}}
\vskip -0.8 true cm
\caption{Logarithm of the optical luminosity $\log L(\nu_R)^{12h}$
  versus redshift $z$ for the observed GRBs updated with the ones
  detected by {\it Swift} (starred dots), for the undetectable simulated
  events (small triangles) and for the observable simulated events
  (empty pentagon).
}
\label{simula}
\end{figure}

In the first case we would expect that the bimodality in the optical afterglow
luminosity is accompanied by some hints of bimodality in other properties,
such as the energetics of the prompt emission or the luminosity distribution
of the X--ray afterglow flux.  However, as already discussed in Nardini et al.
(2006a), there are no convincing evidences of such a behaviour.  Indeed, it
was just this lack of evidence that prompted us to consider the alternative
hypothesis of grey absorption.  In this case, however, we have to assume the
presence of a relatively large amount of grey dust (producing an absorption of
1.5--2 magnitudes) only in a fraction of bursts. Our results are therefore
difficult to understand, implying, in any case, the existence of two,
relatively well separated, GRB optical afterglow families.
Note that infrared observations, 
while surely important and crucial to confirm the existence of the clustering
of the high luminosity burst class and the separation in two luminosity
classes, would not discriminate between the two hypotheses mentioned above.
In fact, since the grey dust is absorbing the observed infrared flux (which
would be optical or UV in the rest frame) by the same amount of the observed
optical one, we could not distinguish between an intrinsic bimodal luminosity
distribution and the presence of grey dust. To this aim it
  will be useful in the future an extended broadband afterglow
  analysis of the optically subluminous events. On the other hand infrared
observations and spectroscopy would not be limited to bursts having $z<5$, and
could therefore give important information on the number of bursts above this
redshift limit, where the different star formation rates greatly differ.  It
will then be possible to directly measure the number of bursts which are
optically dark because they lie at large redshifts.  In Fig. \ref{simula}, we
superposed the simulated events to the plot shown in Fig. \ref{lumlim}. In
this case we simulated two separate Gaussian luminosity functions
characterised by the same width but with different mean values (i.e. 30.65 and
29). The starred dots represent the observed values updated with the {\it
  Swift} GRBs, the small triangles represent the non detectable simulated
events and the empty pentagons represent the observable simulated ones. This
figure shows how such a bimodal optical luminosity function well reproduces
the distribution obtained with the real data.
\\

Our results have important implications for the nature of dark GRBs.
In the pre-Swift era it was possible to infer a redshift for at least
3 Dark GRBs (i.e. 970828, 000210 \& 000214). In Nardini et al. (2006a) we
showed that the X-Ray and Gamma-Ray energetics of GRB 000210 and GRB 000214
are comparable to the optically subluminous ones. The Galactic
absorption corrected R band upper limits of GRB 000210 and of GRB
970828 are deep enough to infer upper limits for their intrinsic
optical luminosities that are much smaller than the observed events
ones (i.e. $\log{L(\nu_R)}<28.7$ at 8.9h after trigger rest frame for
  GRB 000210 and $\log{L(\nu_R)}<28.4$ at 3.2h after trigger rest frame).
Lazzati et al. 2002 excluded that most of the long Dark GRBs
  have no optical detection
because of ``adverse'' observing conditions, since their associated limiting
optical magnitudes are not particularly small (i.e. the corresponding flux
limits are severe).  Therefore it is very likely that dark bursts belong to
the observed optically underluminous family, whose ``existence" has been
statistically proved in this work.  If it is intrinsically faint or affected
by a large achromatic extinction is still to be found.\\

We did our study using a sample of bursts belonging to the pre--{\it Swift}
era.  As stated, we are forced to do that, because of the paucity of {\it
  Swift} events with measured (and published) values of the host galaxy
extinction.  Once we will have enough bursts, we can repeat our analysis with
{\it Swift} bursts, which will shed light on the link between the optical
afterglow properties at relatively late times (12 hours in the rest frame) and
the early afterglow properties.  For the moment, we can only stress that all
{\it Swift} bursts for which the host galaxy extinction has been measured
entirely confirm the observed bimodality in the optical luminosity
distribution.

\section{Summary and conclusions}

In this study we analysed the possible importance of observational selection
effects on the clustering and the bimodality found in the long GRBs optical
afterglow luminosity distribution.

\begin{itemize}
\item We studied the $R$ band upper limits of all the pre--{\it Swift} dark
  GRBs and we showed that they are consistent with the ``typical'' afterglow
  detections.  The distribution of these upper limits extrapolated at 12 h
  after trigger enables the definition of a Telescope Selection Function,
  which can define the probability, for any burst, to be pointed with a
  telescope and exposure time corresponding to some limiting magnitude.
\item Through Montecarlo simulations, we have studied which combinations of
  optical luminosity functions and absorption in the host galaxy can be
  consistent with the observations.  In doing so, we have found that the gap
  between the two observed luminosity distribution is real and it is not the
  result of a small number statistics.
\item If the absorption is chromatic (i.e. a ``standard" one) no unimodal
  intrinsic luminosity distribution agrees with the observed one.  If a good
  fraction of events (but not all) is absorbed by ``grey" (i.e. achromatic)
  dust, then a unimodal luminosity distribution is possible.  An achromatic
  absorption would not be recognised through the standard $A_V^{host}$
  estimate methods.  This can lead to underestimate the intrinsic luminosity
  for a fraction of bursts that could thus appear as members of an
  underluminous family.
\item Dark bursts could then be associated either to an intrinsically
  optically underluminous family, or to those bursts being characterised by a
  relatively large achromatic absorption.  In the first case one should
  explain why, optically, the luminosity distribution is bimodal, while other
  properties are not (e.g. the distribution of the energetics of the prompt
  emission), while in the second case one should explain why a fraction of
  bursts live in a different environment, characterised by grey dust.
\end{itemize}

\section*{Acknowledgements}
We thank a PRIN--INAF 2005 grant for funding.
We would like to thank Annalisa Celotti and Vladimir Avila Rees for
useful discussions and the anonymous referee for the suggestion of a
better statistical check.

\section{Appendix 1 \\Reference for the upper limits of the dark burst} 

GRB970111: Castro-Tirado et al. (1997);
GRB970228: Odewahn et al. (1997);
GRB981220: Pedersen et al. (1999);
GRB990217: Palazzi et al. (1999);             
GRB990506: Vrba et al.(1999);
GRB990527: Pedersen et al. (1999b);
GRB990627: Rol et al. (1999);
GRB990704: Rol et al. (1999b);
GRB990806: Greiner et al. (1999);
GRB990907: Palazzi et al. (1999b);
GRB991014: Uglesich et al. (1999);
GRB991105: Palazzi et al. (1999c);
GRB991106: Jensen et al. (1999);
GRB000115: Gorosabel et al. (2000);
GRB000126: Kjernsmo et al. (2000);
GRB000210: Gorosabel et al. (2000b);
GRB000307: Kemp et al. (2000);
GRB000323: Henden et al. (2000);
GRB000326: Pedersen et al. (2000);
GRB000408: Henden et al. (2000b);
GRB000416: Price et al. (2000);
GRB000424: Uglesich et al. (2000);
GRB000508B: Jensen et al. (2000);
GRB000519: Jensen et al. (2000b);
GRB000528: Palazzi et al. (2000);
GRB000529: Palazzi et al. (2000b);
GRB000607: Masetti et al. (2000);
GRB000615: Stanek et al. (2000);
GRB000616: Bartolini et al. (2000);
GRB000620: Gorosabel et al. (2000c);
GRB000623: Gorosabel et al. (2000d);
GRB000801: Palazzi et al. (2000c);
GRB000812: Masetti et al. (2000b);
GRB000830: Jensen et al. (2000c);
GRB001018: Bloom et al. (2000);
GRB001019: Henden et al. (2000c);
GRB001025: Fynbo et al. (2000);
GRB001105: Castro Ceron et al. (2000);
GRB001109: Greiner et al. (2000);
GRB001120: Price et al. (2000b);
GRB001204: Price et al. (2000c);
GRB001212: Zhu (2000);
GRB010103: Dillon et al. (2001);
GRB010119: Price et al. (2001);
GRB010126: Masetti et al (2001);
GRB010213: Zhu (2001);
GRB010220: Berger et al. (2001);
GRB010324: Oksanen et al (2001);
GRB010326A: Price et al. (2001b);
GRB010326B: Pandey et al. (2001);
GRB010412: Price et al. (2001c);
GRB010629: Halpern et al. (2001);
GRB011019: Komiyama et al. (2001);
GRB011030: Rhoads et al. (2001);
GRB011212: Saracco et al. (2001);
GRB020127: Castro Cerón et al. (2002);
GRB020409: Price et al. (2002);
GRB020418: Gorosabel et al. (2002);
GRB020531: Dullighan et al., (2002);
GRB020603: Castro Cerón et al. (2002b);
GRB020604: Gorosabel et al. (2002b);
GRB020625: Price et al. (2002b);
GRB020812: Ohashi et al. (2002);
GRB020819: Levan et al. (2002);
GRB021008: Castro-Tirado et al. (2002);
GRB021016: Durig et al. (2002);
GRB021112: Schaefer et al. (2002);
GRB021113: Kawabata et al. (2002);
GRB021201: Garnavich et al. (2002);
GRB021204: Ishiguro et al. (2002);
GRB021206: Pedersen et al. (2003);
GRB021219: Castro-Tirado et al. (2002b);
GRB030204: Nysewander et al. (2003);
GRB030320: Gal-Yam et al. (2003);
GRB030413: Schaefer et al. (2003);
GRB030414: Lipunov et al. (2003);
GRB030416: Henden et al. (2003);
GRB030501: Klotz et al. (2003);
GRB030823: Fox et al. (2003);
GRB030824: Fox et al. (2003b);
GRB030913: Henden et al (2003);
GRB031026: Chen et al. (2003);
GRB031111: Silvey et a. (2003);
GRB040223: Gomboc et al. (2004);
GRB040228: Sarugaku et al. (2004);
GRB040403: de Ugarte et al. (2004);
GRB040624: Fugazza et al., (2004);
GRB040701: de Ugarte et al. (2004b);
GRB040810: Price et al. (2004);
GRB040812: Cobb et al. (2004);
GRB040825A: Jensen et al. (2004);
GRB040825B: Gorosabel et al. (2004);
GRB041015: Isogai et al. (2004);
GRB041016: Kuroda et al. (2004);
GRB041211: Monfardini et al. (2004);

\end{document}